\documentstyle[aps,pre,epsf,multicol]{revtex}
\begin{document}
\draft
\title{Directed Ising type dynamic preroughening transition
in one dimensional interfaces}
\author{Jae Dong Noh, Hyunggyu Park
\thanks{Permanent Address: Department
of Physics, Inha University, Inchon 402-751, Korea.},
and Marcel den Nijs}
\address{Department of Physics, University of Washington,
P.O. Box 351560, Seattle, Washington 98195-1560, U.S.A.}

\date{\today}
\maketitle

\begin{abstract}
We present a realization of directed Ising (DI) type dynamic
absorbing state phase transitions
in the context of one-dimensional interfaces,
such as the relaxation of a step on a vicinal surface.
Under the restriction that particle  deposition and evaporation
can only take place near existing kinks,
the interface relaxes into one of three steady states:
rough, perfectly ordered flat (OF) without kinks, or
disordered flat (DOF) with randomly placed kinks but in perfect up-down
alternating order.
A DI type dynamic preroughening transition takes place between the
OF and DOF phases.
At this critical point
the asymptotic time evolution is controlled not only by the DI exponents
but also by the initial condition.
Information about the correlations in the initial state persists
and changes the critical exponents.

\end{abstract}


\begin{multicols}{2}
\narrowtext
\section{Introduction}\label{sec1}

Absorbing type dynamic phase transitions are the focus of
extensive research~\cite{Marro,Liggett85,Grassberger_Torre79,Jensen93,Takayasu_Tretyakov92,Grassberger_Krause_Twer84_Grassberger89,Menyhard94,Kim_Park94,Jensen94,Park_Park95,Hinrichsen97,Hwang_Park98}.
These transitions occur in dynamic processes with trapped states.
At the absorbing side of the phase transition,
the system evolves into one specific microscopic state,
a so-called absorbing state, out of which it can not escape.
At the active side of the phase transition,
the system manages to avoid such traps.
An ensemble  has a finite probability to stay alive.

Two distinct types of  absorbing phase
transitions have been identified in one dimension:
the directed percolation~(DP)
and directed Ising~(DI) universality class.
DP type dynamic critical behavior has been found in,
e.g.,~Schl\"{o}gl's first model for contact
processes~\cite{Liggett85,Grassberger_Torre79},
pair contact processes~\cite{Jensen93}, and
branching annihilating random walks~(BAW)
with an odd-number of offspring~\cite{Takayasu_Tretyakov92}.
DI type dynamic critical behavior has been found  in,
e.g.,~probabilistic cellular
automata~\cite{Grassberger_Krause_Twer84_Grassberger89},
nonequilibrium kinetic Ising type models~\cite{Menyhard94},
interacting monomer-dimer models~\cite{Kim_Park94},
and BAW models with an even-number  of
offspring~\cite{Takayasu_Tretyakov92,Jensen94}.
These models describe a wide range of phenomena,
in particular epidemic spreading and catalytic chemical reactions.

In analogy with equilibrium phase transitions,
it is believed that dynamic universality classes
are determined by the degeneracy and symmetries of the absorbing states
and by the symmetry properties of interfaces (domain walls) between
them~\cite{Park_Park95,Hinrichsen97,Hwang_Park98}.
DP type transitions involve typically  only a single absorbing state
or a set of absorbing states
with one of them dynamically more prominent in a coarse grained sense
than the others~\cite{Jensen93,Park_Park95,Hwang_Park98}.
DI type transitions involve
two equivalent adsorbing states or two equivalent classes of
absorbing states~\cite{Hwang_Park_un}.

Counting the number of absorbing states is not sufficient.
The degeneracy of the absorbing states can be obscured by the
formulation of the dynamic rule.
For example, in BAW dynamics involving only one species of particles,
the ``empty" absorbing state seems not to be degenerate,
but the transition belongs to the DI universality class
if the dynamics conserves the particle number modulo
2~\cite{Takayasu_Tretyakov92,Jensen94}.
In that case particles can be reinterpreted as domain walls
and we can color the domains at opposite sides of the walls alternatingly
with two colors. One of the two colors dies out in the adsorbing state.
This is reminiscent of the domain wall formulation of the equilibrium
Ising model, where the existence of two coexisting phases is also obscured.

One appealing scheme of classifying absorbing phase transitions is by
association with conventional equilibrium phase transitions.
For example, consider the one-dimensional
Ising model with single spin-flip dynamics.
In terms of domain walls this involves the following processes:
spontaneous creation of domain wall pairs $0 \to 2A$ with probability
$p_{sp}$ (the Boltzmann weight),
hopping of single domain walls with probability $p_h=1$, and
pair annihilation $2A \to 0$ with probability $p_{a}=1$.
The stationary state of this model is the equilibrium state, which is
obviously  ``disordered" (active) in one dimension.
However, the model evolves always into one of the two perfectly
ordered (absorbing) states
if we disallow the spontaneous creation of domain walls, $p_{sp}=0$,
because then the domain wall density can only decrease.
The DI transition comes into play when  branching processes
are allowed; i.e., the creation of domain wall pairs in the vicinity of
existing domain walls,  $A \to 3A$.
Branching can keep the system alive, while the two perfectly ordered
Ising ground states still remain absorbing states.
The resulting model is known as the BAW model with two offspring.

The same line of reasoning associates a distinct
absorbing type  dynamic universality class  with
each conventional  equilibrium universality class,
like  the familiar $q$-state Potts and  $q$-state clock universality
classes. However, none of the models studied thus far
with higher symmetries than the Ising model
has an absorbing phase transition.
Dynamic processes  with more than 2 equivalent absorbing states
$(q\ge 3)$ appear to be always active~\cite{Hinrichsen97,Noh}.
Cardy and T\"{a}uber provide a possible explanation for this
\cite{Cardy_Tauber96}.
Additional systematic numerical studies
are needed to settle this issue.

In this paper we introduce an application of
absorbing phase transitions to the dynamics of
one-dimensional solid-gas interfaces.
We find a one-dimensional dynamic preroughening (PR) transition from an
ordered flat (absorbing) phase to a disordered flat (active) phase.
Consider the one-dimensional restricted solid-on-solid (RSOS) model
description of crystal surfaces. The height difference between
nearest neighboring columns of particles is 0 or $\pm 1$,
i.e.,~only steps (kinks) with single atomic heights are allowed.
A conventional dynamic Monte Carlo type equilibration process
where single particles adsorb on or desorb from the surface
with equal probability leads to a rough equilibrium surface with
Edwards-Wilkinson (EW) type~\cite{EW} scaling properties.
In analogy with the above directed Ising model discussion,
we transform the perfectly ordered flat state
into an absorbing state  by  disallowing adsorption and desorption of
particles at flat segments of the surface (spontaneous creation of step pairs).
The only processes allowed are:
hopping/pair annihilation of steps by
adsorbing or desorbing a particle at step edges
and branching of steps by adsorption or desorption at
the next nearest neighbour sites near existing steps.
The branching processes fall into two classes:
the ones that preserve local
flatness and those that create local roughness.
We give them independent probabilities.

This model describes the evolution of the one dimensional interface
from any initial random rough
configuration to three types of stationary states:
a rough phase with dynamic exponent $z=2$ (EW type),
a perfectly ordered flat (OF) absorbing phase without steps,
and an active disordered flat (DOF) phase.
DOF phases are well known in equilibrium surfaces.
They represent step liquids where the surface remains flat due to
long-range step-up step-down alternating order
but steps are placed randomly~\cite{Rommelse_denNijs87,denNijs94}.
Our dynamic DOF stationary state has no defects, i.e.,
the up-down alternating order is perfect and the
the steps are placed randomly.

Although the dynamic DOF phase is active,
it has an absorbing state type property.
The OF absorbing state and the DOF active phase at the other side of
the transition have in common that
both  ``ground states" lack thermodynamic defects
(both are at their fixed point in the sense of renormalization theory).

Equilibrium preroughening transitions belong to the Ashkin-Teller
universality class~\cite{Rommelse_denNijs87,denNijs94}.
They involve two coupled Ising type order parameters.
They are non-zero at opposite sides of the transition.
A perfectly DOF initial configuration has a half integer average surface
height
$n+\frac{1}{2}$  and evolves at the OF side of the PR transition
into a perfectly flat ordered absorbing state with $n$ or $n+1$ integer
surface height.
This OF-type two-fold degeneracy
is the conventional spontaneous symmetry breaking associated with
DI type transitions.
However, an Ising type degeneracy exists also at the DOF side of
the PR transition.
An initial state with an integer average surface height $n$,
evolves into a stationary DOF state with surface height
$n+\frac{1}{2}$ or $n-\frac{1}{2}$.
We call this the DOF-type degeneracy.
In analogy with equilibrium transitions
one might expect therefore that the dynamic PR transition  be
described by 2 coupled DI transitions.
This turns out not to be the case.

The dynamic PR transition belongs to the (single) DI universality class.
The two types of spontaneous symmetry breaking  are not on an equal footing.
A perfect DOF initial state can decay into
the OF state, but a perfect OF state is frozen forever
(even at the DOF side of the PR transition).
The DOF type degeneracy does not create additional DI critical fluctuations.
But it still affects the scaling behaviour.
Configurations with perfect DOF order form an invariant subspace
(which includes the OF ordered state).
As shown below, this leads to a peculiar initial configuration dependence 
of the asymptotic decay of the step densities
(the kinks in the one-dimensional interface).

In section 2, we introduce our model in detail.
It is instructive to interpret it
not only as a model for surface relaxation, but also for surface catalysis.
In its latter reincarnation,
the process is a two-species generalization of
the BAW model with two offspring.
The up and down steps (kinks) represent two species of particles, $A$ and $B$.

In section 3 we limit ourselves to configurations
with perfect $ABAB$ alternating order.
These DOF type configurations form a dynamical invariant subspace.
The numerical results presented in section 3
confirm that our PR transition belongs to the DI universality class.

In section 4, we discuss the crossover scaling
properties of the DI critical point into the rough phase.
The rough phase has conventional EW type scaling properties.
Recall that we disallow spontaneous adsorption and desorption from
flat surface segments;
surface roughness can only be created and maintained by branching.
Apparently, this restriction does not alter the scaling properties of the
rough phase.

The DOF-type Ising degeneracy shows up
in the evolution of arbitrary initial states.
The kink densities scale in time in an unusual way,
not only at the PR transition, but  also
everywhere in the OF and DOF phases.
They depend strongly on the initial conditions;
whether the initial configuration is flat or rough,
and on the correlations in such initial rough states.
Conventional wisdom tells us that
the long time scaling of dynamic processes
depends only on the dynamic  exponent and
the  stationary state exponents of a process.
In our case, critical exponents vary with the initial conditions.
We present an analytical scaling theory for
this in section 5 and numerical results in section 6.

\section{The Model}\label{sec2}

Consider a one-dimensional lattice.
Each site is vacant or occupied by at most one $A$ or one $B$ type particle.
In the surface catalysis interpretation,
$A$ and $B$ represent two types of particles.
In the surface formulation, they  represent up and down steps.
Configurations evolve in time according to the following dynamic rule.
First choose a site at random. If the site is empty, nothing happens.
If the site is occupied, the particle
can hop to its nearest-neighbor sites with probability $p$.
If it lands on top of an existing particle of the opposite kind,
the $AB$ pair annihilates immediately.
The move is rejected if the particle would land on top of a particle
of the same type.

\begin{figure}
\centerline{\epsfxsize=75mm \epsfbox{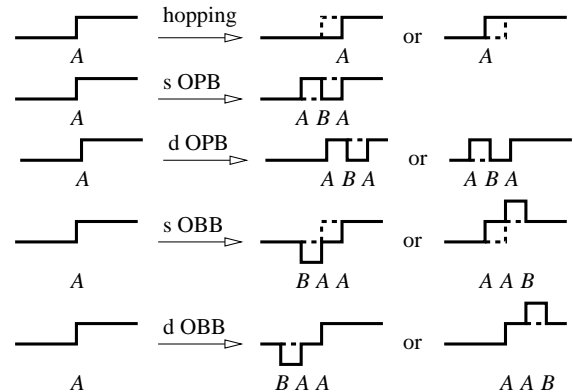}}
\caption{
Step hopping and branching processes of an up step
(an $A$ type particle).
The letters $s$ and $d$ represent static and dynamic type branching,
respectively.
The dashed lines denote the surface before each event.
Dynamic events near B particles are identical by mirror symmetry.}
\label{rule}
\end{figure}

Besides hopping, each  particle can also
branch into three particles by the creation of an $AB$ pair.
Branching comes in two distinct flavours;
the one that preserves local $ABAB$ order and the one that breaks it.
Order preserving branching (OPB)
creates local DOF step-up  step-down order in the surface.
Order breaking branching (OBB) creates local roughness (see Fig.~\ref{rule}).
These branching processes occur with probability $q$ and $r$, respectively.
Branching could lead into double occupancy of a site.
$AB$ pairs at the same site annihilate immediately.
The branching event is rejected if it would result in two particles
of the same type at any site.
We require that the chosen particle must attempt to hop
or branch. This implies that $p+q+r=1$.
$\tilde{q}\equiv q/p$ and $\tilde{r}\equiv r/p$ are our choices
for independent parameters.

We need to distinguish also between so-called dynamic and static branching.
The two new particles, $A$ and $B$, created by each branching process
may be placed in two different ways.
The center of mass is stationary (static branching)
or moves (dynamic branching)~\cite{Kwon_Park95}.
In the latter the new particles are placed both to the left or
both to the right of the parent particle with equal probability.
In the interface representation of the model,
dynamic branching corresponds to single particle adsorption/desorption.
Static branching represents two-particle events (see Fig.~\ref{rule}).
So dynamic branching is more natural.
These details do not affect the universality of the phase transition.
But they can change dramatically the location of the critical
point\cite{Kwon_Park95}.
It is well known that in the BAW model with two offspring the stationary
state is always absorbing.
Static branching does not create enough activity to destabilize the
absorbing phase~\cite{Sud&Tak_92}.
We applied both type of branching in our simulations.
The universal scaling properties do not change.
Here only our results for dynamic branching are presented.

Each particle configuration of the surface catalysis model
maps onto a surface height configuration of the RSOS type surface growth model.
The $A$ and $B$ particles represent up and down steps of unit height.
The above dynamic rules for the $A$ and $B$ type particles
translate into placement or removal of a single surface atom
near existing steps (Fig.~\ref{rule}).
Adsorptions and desorptions have equal probability.
The surface does not advance nor retreat on average.
Flat segments of the surface are inactive,
which makes the perfectly ordered flat state  an absorbing state.

\begin{figure}
\centerline{\epsfxsize=7cm \epsfbox{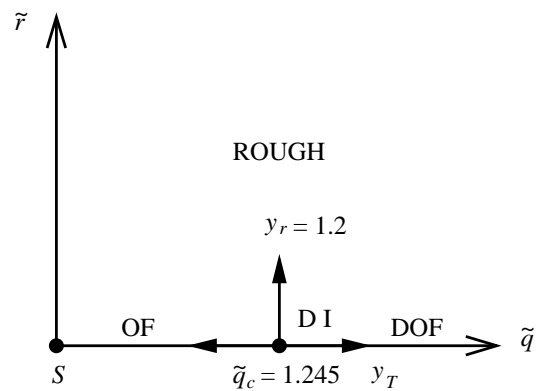}}
\caption{
Phase diagram of our model.
$\tilde r = r/p$ is the probability for OBB type branching
(creating local surface roughness) and
$\tilde q = q/p$ is the probability for OPB type branching
(creating disordered flat type local order), relative to the hopping
probability $p$. $y_r$ and $y_T\equiv 1/\nu_\perp$
are the scaling dimensions of the OBB and OPB operators, respectively.
}
\label{phasediagram}
\end{figure}

The structure of the phase diagram is shown in Fig.~\ref{phasediagram}.
The numerical details will be presented in section 3 and 4.
Here we want to point out the general features.

OPB type branching preserves the average surface height, while
OBB type branching creates local roughness.
This suggests two distinct types of active phases.
In the region of the phase diagram where
OPB type branching is dominant, the steps prefer up-down alternating order.
The surface likes to remain flat on average with randomly placed steps.
This is a one-dimensional dynamical version of the DOF phase
known from two-dimensional equilibrium
crystal surfaces~\cite{Rommelse_denNijs87,denNijs94}.
In the region of the phase diagram where OBB type branching is dominant,
the active phase lacks step up-down alternating order and is therefore
probably  rough.

In the presence of OBB type branching ($\tilde{r}>0$),
the stationary surface state turns out to be always rough
(see Fig.~\ref{phasediagram}).
Along the $\tilde{r}=0$ line, the surface is flat;
in  the  OF absorbing phase for $\tilde q < \tilde q_c$
and the DOF active phase  for $\tilde q > \tilde q_c$.
Spontaneous order is difficult to maintain in one dimension.
This explains why the DOF phase is limited to the $\tilde{r}=0$ subspace.
It would have been nice, but a surprise if the OF phase extended into
$\tilde r>0$.
Equilibrium roughening phase transitions can be viewed as $q\to \infty$ limits
of the $q$-state clock model. Absorbing phase transitions
with  $q \ge 3$ are unknown and may not exist at all
as mentioned in section 1~\cite{explain0}.

The transition point $\tilde q_c$  in Fig.~\ref{phasediagram}
represents a one-dimensional dynamic analogue of equilibrium
preroughening transitions.
The disordered flat phase is active (a step liquid phase),
but maintains perfect  step-up step-down alternating order.
$A$ and $B$ type particles move around like in a liquid, but
remain perfectly  $ABAB$ ordered.
The distances between neighboring $A$ and  $B$ particles are randomly
distributed.

To determine the scaling  properties of this transition (section 3)
we measure the step (particle) density $\rho (t)$ as a function of time $t$,
which is the sum of the two single particle densities, $\rho_A (t) + \rho_B
(t) $.
The difference between them,
$\rho_A (t) - \rho_B (t) $, describes the global tilt of the surface and
is preserved by the dynamic rule.
The step density $\rho$ in the stationary state vanishes in the absorbing 
phase and remains non-zero in the two active (rough and DOF) phases.

We monitor also the density of $AB$ pairs of particles,
$\rho_{AB}(t)$.
The $A$ and $B$ particles in each $AB$ pair do not have to be immediately
adjacent to each other.
They can be separated by a stretch of empty space of arbitrary length.
Similarly, $\rho_{AA}(t)$ and $\rho_{BB}(t)$
are the densities of $AA$ and $BB$ pairs.
It is easy to show that $\rho_A(t)=\rho_{AA}(t)+\frac{1}{2}\rho_{AB}(t)$
and $\rho_B(t) = \rho_{BB}(t)+\frac{1}{2}\rho_{AB}(t)$.
$\rho_{AA} (t) - \rho_{BB} (t)$ is preserved by the
dynamics and equal to zero for nontilted surface configurations.
$\rho_{AA}$ distinguishes between the DOF and the rough active phases.
In the DOF stationary state, the system is active with perfect $AB$
alternating order, i.e.,~$\rho \neq 0$ and  $\rho_{AA} = 0$.
In the rough stationary state, the $AB$ alternating order is broken,
i.e.,~$\rho\neq 0$ and $\rho_{AA} \neq 0$.

\section{The Directed Ising Dynamic Preroughening Transition}\label{sec3}

In this section we present numerical evidence for the DI nature
of the dynamic preroughening  transition.
Consider the  $\tilde r=0$ line of the  phase diagram
(Fig.~\ref{phasediagram}) where OBB type branching is excluded.
Here, the configurations with perfect $ABAB$ alternating DOF type order form a
dynamically  invariant subspace.

In this subspace our model is almost identical to the
BAW model with 2 offspring and dynamic branching~\cite{Kwon_Park95}.
There, each site may be empty or occupied by a particle of a single species.
Those particles can hop to a nearest neighbor site with probability $p$ or
create two offspring on the nearest and next-nearest neighbor sites to the
left or to the right with probability $q=1-p$ (dynamic branching).
Two particles annihilate immediately when they happen to land on the same site.
This BAW model exhibits an absorbing phase
transition at $p= 0.5105(7)$, which belongs to the DI universality
class~\cite{Kwon_Park95}.

There is an exact two-on-one mapping between the configurations
in our model and those in the above BAW model.
Simply label the particles in the latter
as  $A$ and $B$ (or $B$ and $A$) alternatingly.
The dynamic processes for the two models are virtually the same,
except for one detail.
In the BAW model, without $A$ and $B$ labels, particles can always
annihilate when they land on the same site.
In our model only $A$ and $B$ pairs can annihilate, and
$AA$ and $BB$ pairs repel each other.
Hopping events are not affected by this, but
OPB type branching processes are.
Consider a configuration with an  isolated $AB$ nearest pair, like
$000AB000$ where
$0$ represents a vacant site.
Suppose that the $A$ particle is chosen to branch a pair of
$BA$ particles to the right.
In our model, this attempt is rejected because it would result
in two $B$ particles on a single site.
In the the BAW model this attempt is accepted,
and results in the annihilation of two particles.
This difference between the two models does not affect the mod 2
conservation of
the total particle number. So we expect a DI-type absorbing phase transition
along the $\tilde{r}=0$ line, but the OF phase must be more stable.

The mapping between configurations of the BAW model and our model
is lost outside  the  configurational subspace  with perfect $ABAB$ order.
The  DOF subspace is an attractor, however.
It contains the global stationary state for arbitrary initial configurations.
Consider an arbitrary configuration, i.e., one with $AA$ and $BB$ pairs.
For $\tilde r =0$ there is no mechanism to
increase their total numbers, $N_{AA}$  and $N_{BB}$.
Hopping and OPB type branching decrease $N_{AA}$ and $N_{BB}$,
by annihilation of $AB$ pairs but never increase them.
For example, consider a configuration like $0A0AB0B0$.
Hopping of the central $A$ particle to the right induces the annihilation
of an $AB$ pair
and the configuration becomes $0A0000B0$.
$N_{AA}$  and $N_{BB}$  decrease by one.
The density of these pairs, $\rho_{AA} =\rho_{BB}$, decreases monotonically
in time.
In fact, they decay algebraically to zero along the entire
$\tilde{r}=0$ axis (see section 5 and 6).
The active steady state involves only configurations with perfect
$ABAB$ alternating order.

To locate the critical point we perform defect dynamics
type Monte Carlo simulations, in which one starts with a single
nearest-neighbor
pair of $AB$ particles at the center of an empty lattice.
Obviously, this initial configuration belongs to the DOF type invariant
subspace.
Time increments by one unit after $L$ single site updates
(one Monte Carlo step) with $L$ the lattice size.
We measure the survival probability $P(t)$ (the probability that
the system is still active at time $t$) and the number of kinks (particles)
$N(t)$
averaged over $5\times 10^5$ independent runs up to 5000 time steps.

\begin{figure}
\centerline{\epsfxsize=85mm \epsfbox{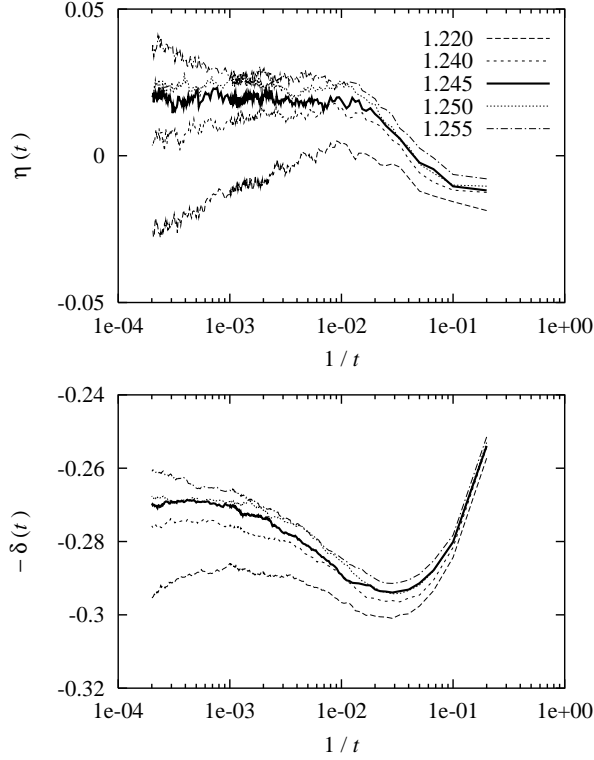}}
\caption{
Semi-log plots of the effective exponents $\eta$ and
$\delta$ versus $1/t$ for several values of $\tilde q$
close to criticality.
The data at our best estimate for the critical
point $\tilde q_c$ are highlighted as thick lines.}
\label{defect}
\end{figure}

At criticality the long time limits of these two quantities are governed by
power laws
with critical exponents $\delta$ and $\eta$ as
$ P(t) \sim t^{-\delta} $ and $N(t) \sim t^{\eta}$~\cite{Grassberger_Torre79}.
Precise estimates for the
critical point and the critical exponents are obtained from a finite
time analysis of
\begin{eqnarray}
-\delta(t) &=& \frac{\ln[P(t)/P(t/m)]}{\ln m} \\
\eta(t) &=& \frac{\ln[N(t) /
            N(t/m)]}{\ln m} .
\end{eqnarray}
In Fig.~\ref{defect}, we plot these effective exponents against
$1/t$ with $m=5$ for several values of $\tilde{q}$.
These plots bend up or down in time except at criticality.
This leads to an estimate of the critical point,
$\tilde{q}_c=1.245(5)$~\cite{explain1}, and also for the critical exponents,
\begin{equation}
\label{delta&eta}
\delta=0.270(5) \quad , \quad \eta=0.015(10).
\end{equation}
These values are slightly different from the standard DI values;
$\delta = 0.285$ and $\eta=0.000$. However, their sum
is related to the steady-state properties via the generalized
hyperscaling relation~\cite{Mendes94}
and is in excellent agreement with the value of the DI universality class.

Further evidence of the DI nature of the transition is obtained by monitoring
the time evolution of the step density $\rho(\delta \tilde{q},t,L)$
in stationary Monte Carlo simulations on a finite lattice of size $L$
with $\delta \tilde{q} \equiv \tilde{q}-\tilde{q}_c$.
We start with a random configuration inside the invariant subspace
with initial kink densities $\rho_A (0) =\rho_B (0) =1/4$
and  periodic boundary conditions.
The step density $\rho$ averaged over survived samples only
should obey the  scaling relation
\begin{equation}\label{rho_s}
\rho(\delta\tilde{q},t,L) = b^{-\beta / \nu_\perp} \rho(b^{1/\nu_\perp}
\delta\tilde{q},b^{-z} t,b^{-1} L).
\end{equation}
$\beta$ is the order parameter exponent, $\nu_\perp$ the
correlation length exponent in the spatial direction,
$z=\nu_\parallel/\nu_\perp$ the
dynamic exponent~\cite{explain2},  $\nu_\parallel$ the correlation length
(relaxation time) exponent in the
time direction, and $b$ an arbitrary scale factor.

This scaling relation determines all scaling properties of $\rho$.
At the critical point, $\rho$ decays algebraically;
$\rho \sim t^{-\beta/\nu_\parallel}$
for $t \ll \tau_L$ with $\tau_L$ a characteristic time scale which diverges
as $L^z$. $\rho$ scales at
$\rho\sim L^{-\beta/\nu_\perp}$ for $t\gg \tau_L$.

\begin{figure}
\centerline{\epsfxsize=85mm \epsfbox{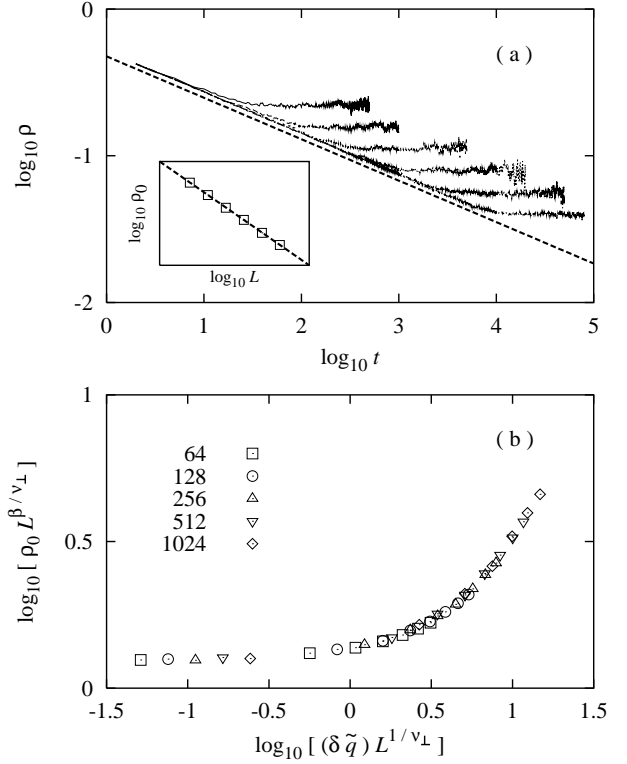}}
\caption{
(a) Decay of the step density at $\tilde q =\tilde{q}_c$.
Each curve corresponds to $L=2^5,\ldots,2^{10}$ from top to bottom.
From the asymptotic slope of the curves we obtain
$\beta/\nu_\parallel=0.282(5)$. The broken line is of slope 0.282.
The inset shows the finite size scaling of the saturated step density.
From the slope we obtain $\beta/\nu_\perp = 0.497(5)$.
The broken line has slope 0.497.
(b) Scaling plot for $\log_{10} \left[\rho_0 L^{\beta/\nu_\perp}\right]$
against $\log_{10} \left[\delta\tilde{q} L^{1/\nu_\perp}\right]$. Using the
values of $\tilde{q}_c$ from defect dynamics simulations and
$\beta/\nu_\perp$ in (a), the best data collapse is obtained with
$\nu_\perp=1.78$.}
\label{static}
\end{figure}

Log-log plots of $\rho(0,t,L)$ versus $t$
are shown in Fig.~\ref{static}(a) at criticality for $L=2^5,\ldots,2^{10}$.
The step densities are averaged over $10^3$ independent Monte Carlo runs
at $\tilde q = \tilde q_c = 1.245$.
The inset shows the saturated values of the step density, denoted by $\rho_0$,
as function of $L$.
From these plots we obtain
\begin{equation}
\label{beta&nu}
\frac{\beta}{\nu_\parallel} = 0.282(5) , \quad
\frac{\beta}{\nu_\perp} = 0.497(5) ,
\end{equation}
and $z=1.76(1)$ from their ratio.
These results are in excellent agreement with
those of the DI universality class.

The above analysis at the critical point gives only ratios of critical
exponents.
Their bare values can be extracted from the off-critical behavior of $\rho$.
The saturated value of the step density follows the scaling form
$\rho_{0}(\delta\tilde{q},L)=
L^{-\beta/\nu_\perp} {\cal G}(\delta\tilde{q}\ L^{1/\nu_\perp})$,
according to Eq.~(\ref{rho_s}).
The scaling function ${\cal G}(x)$ becomes a constant in the $x \to 0$ limit
and scales as ${\cal G}\sim x^\beta$ in the $x \to \infty$ limit,
because $\rho_{0} \sim (\delta \tilde{q})^\beta$ in the $L\rightarrow
\infty$ limit.
In Fig.~\ref{static}(b), we show the log-log plots of
$\rho_0 L^{\beta/\nu_\perp}$ against
$\delta\tilde{q}\ L^{1/\nu_\perp}$. Using the values of
$\tilde{q}_c$ obtained from the defect dynamics simulations and
$\beta/\nu_\perp$ from
Eq.~(\ref{beta&nu}), we find that the data are best collapsed with
$\nu_\perp=1.78(5)$. Combining this value with those in
Eq.~(\ref{beta&nu}), we obtain the critical exponents
\begin{equation}\label{exponents}
\beta = 0.88(3), \quad \nu_\perp = 1.78(5), \quad \nu_\parallel = 3.12(5) \ .
\end{equation}
These values agree well with the DI values and satisfy the generalized
hyperscaling
relation~\cite{Mendes94}, $\delta+\eta=(\nu_\perp-\beta)/\nu_\parallel$
very well.
We conclude that our model has an absorbing type
preroughening transition between the ordered flat and the DOF phase, which
belongs to the DI universality class.

\section{Crossover into the Rough Phase}\label{sec 4}

\begin{figure}
\centerline{\epsfxsize=85mm \epsfbox{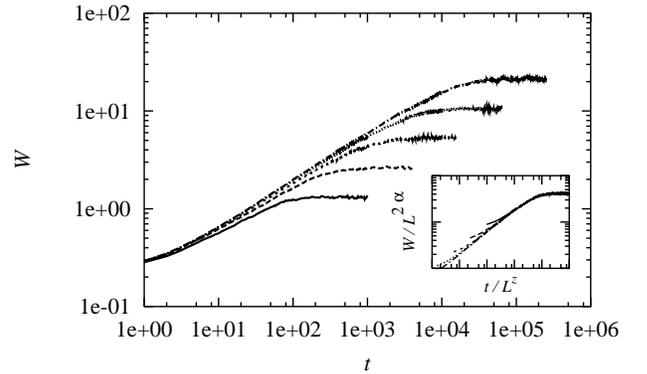}}
\caption{The surface width in the rough phase at $\tilde{q}=\tilde{r}=1.0$.
The curves correspond to system sizes $L=2^5,\ldots,2^{9}$.
The width obeys the scaling form
$W = L^{2\alpha}f(t/L^{\alpha/\beta})$ with the Edwards-Wilkinson
exponents $\alpha=1/2$, $\beta=1/4$, and $z=\alpha/\beta=2$ as shown in the
inset.}
\label{width}
\end{figure}

The PR phase transition is unstable with respect to
OBB type branching, for all $\tilde{r}>0$.
We perform defect dynamics simulations at various values of
$\tilde r$ and $\tilde q$.
The plots for the effective exponents, $\delta(t)$ and $\eta(t)$,
show only upward curvature.
This indicates that the system is always active for all $\tilde r > 0$.
The size of the active region $R(t)$ averaged over survived samples
confirms this.
It grows linearly in time so the spreading velocity of the active region is
finite
in the long time limit.
In this active phase, the densities of $AA$ and $BB$ pairs, $\rho_{AA}$ and
$\rho_{BB}$, are non-zero.
This suggests that the steady-state surface is rough.
We measure the surface width $W(t,L)$
\begin{equation}
W(t,L) = \left\langle \frac{1}{L} \sum_i^L \left(h_i(t)-\bar{h}(t)\right)^2
\right\rangle
\end{equation}
in Monte Carlo simulations on a finite lattice of size $L$.
$h_i (t)$ is the height of the surface at site $i$,
$\bar{h}(t)$ the average height,
and $\langle\cdots\rangle$ the average over survived samples.
In Fig.~\ref{width}, we illustrate the typical behaviour of $W(t,L)$
by the evolution of the surface at $\tilde{q}=\tilde{r}=1.0$,
using as initial condition $\cdots 0A0B0 \cdots$ with system size
$L=2^5,\ldots,2^{9}$ and taking the average over $10^3$ runs.
The data satisfy the dynamic scaling form
$W(t,L) = L^{2\alpha} f(t/L^{\alpha/\beta})$ with the Edwards-Wilkinson
roughness exponent $\alpha = 1/2$ and the growth exponent $\beta=1/4$.
The active dynamic rough phase has the same scaling properties as the
conventional
equilibrium rough phase in one dimension.

The crossover exponent $y_r$ at the DI critical point into the $\tilde r$
direction
must be relevant.
$y_r$ is the scaling dimension of the OBB type branching operator
similar to $y_T = 1/\nu_{\perp}$ which is the scaling dimension of
the OPB type branching operator.
$y_r$ is potentially an independent DI critical exponent
(see Fig.~\ref{phasediagram}).
We obtain $y_r$ numerically by measuring the size of the active
region $R$ at various values of $\tilde r$ along the $\tilde q=\tilde q_c$
line
in defect dynamics type simulations.
Consider the scaling relation
$R(t,\tilde{r}) = b R (b^{-z_{\rm{DI}}} t,b^{y_{r}} \tilde{r})$
with $z_{\rm{DI}}=\nu_{\parallel}/\nu_{\perp}$ the DI dynamic exponent,
and $b$ an arbitrary scale factor.
For  $b =t^{1/z_{\rm DI}}$, $R$ takes the form
$R(t,\tilde{r}) = t^{1/z_{\rm{DI}}}{\cal F}(\tilde{r} \ t^{y_r /z_{\rm{DI}}})$.
The scaling function ${\cal F}(x)$ becomes a constant
in the $x \to 0$ limit,  because $R \sim
t^{1/z_{\rm{DI}}}$ at the DI transition point.
In the $x \to\infty$ limit,
$\cal F$ must scale as
${\cal F}\sim x^{(z_{\rm{DI}}-1)/y_r}$,
because $R$ grows linearly in time in the long time limit for $\tilde{r}>0$.
Therefore the asymptotic value of the spreading velocity of the active region,
$v\equiv \lim_{t\rightarrow \infty} R/t$,
scales as $v \sim \tilde{r}^\kappa$ with
$\kappa = (z_{\rm{DI}}-1)/y_r$.

\begin{figure}
\centerline{\epsfxsize=85mm \epsfbox{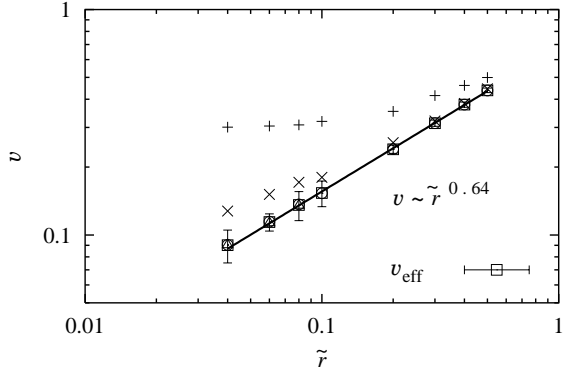}}
\caption{Scaling of the spreading velocity $v$ of the active region.
Different symbols
are used for $v_1~(+)$, $v_2~(\times)$, $v_3~(\circ)$,
$v_4~(\bigtriangleup)$, and $v_{\rm{eff.}}$~($\Box$).
The $v_{k\geq 3}$ merge into $v_{\rm{eff.}}$, which confirms
that $v_{\rm{eff.}}$ is the asymptotic value for $v$.}
\label{crossover_scaling}
\end{figure}

In our simulations, we measure $R$ up to times $t=2\times 10^4$
and average over $2\times 10^3 \sim 5\times 10^3$ samples.
The spreading velocity $v$ is extracted in two ways.
First, we fit $R(t)$ to the form $a_k+v_k t$ in
the time interval $10^k \leq t \le 10^{k+1}$~($k=1,2,3,4$).
$v_k$ will converge to $v$ in the asymptotic regime.
Next, we define  an effective velocity
$v_{\rm{eff}}(t) = (R(t+\Delta t)-R(t))/\Delta t$ with $\Delta t = 500$,
whose saturated value in the asymptotic regime is denoted by $v_{\rm{eff}}$.
Fluctuations around the saturated value give an
estimate of the statistical errors.
In Fig.~\ref{crossover_scaling}, we plot $v_k$ and
$v_{\rm{eff}}$ for several values of $\tilde{r}$.
The estimates for $v_k$ with $k\ge 3$  merge into $v_{\rm{eff}}$.
This confirms that $v_{\rm{eff}}$ is the asymptotic value of the
spreading velocity $v$.
From a power-law fit we estimate $\kappa = 0.64(1)$
and hence obtain the value of the crossover exponent, $y_r = 1.2(1)$.

\section{Density Decay Dynamics}\label{sec 5}

In section 3, we demonstrated the DI nature of the absorbing phase
transition along $\tilde r=0$.
The active stationary state has perfect $ABAB$ alternating order.
The configurations with this perfect DOF type order form an invariant subspace.
This subspace is an attractor, because
the density of $AA$ and $BB$ pairs, $\rho_{AA}=\rho_{BB}$,
never increases in time  in the absence of OBB type branching,
along the $\tilde{r}=0$ line.
In the conventional picture all asymptotic dynamic time scales
depend only on the dynamic exponent $z$ and
the stationary state exponents of the DI transition.
Surprisingly, the  latter is not true in our model.
The step density
\begin{equation}
\rho\sim t^{-\theta}
\end{equation}
and the pair densities
\begin{equation}
\rho_{AA}=\rho_{BB}\sim t^{-\theta_{AA}} \quad , \quad
\rho_{AB}\sim t^{-\theta_{AB}} \quad
\end{equation}
decay in the large time limit
with exponents that depend on the initial condition.
Their values depend on whether the initial state
has perfect $AB$ alternating order or is rough, and whether this roughness
is random or correlated.

These densities decay as power laws everywhere along the $\tilde r =0$ line,
with different exponents in the different phases.
In this section we review first previously known results at point $S$
of the phase diagram, see Fig.~\ref{phasediagram}, and then generalize
those results to the entire $\tilde r =0$ line.

\subsection{diffusion-limited pair annihilation: $\tilde{q}=0$}

At point  $S$ ($\tilde r=\tilde q=0$)
all branching processes are disabled and the
dynamics describes chemical reactions with diffusion-limited
pair annihilation $A+B\rightarrow 0$.
This process has been studied
extensively~\cite{Bramson_Lebowitz88,Toussaint_Wilczek83,Kang_Redner84,Leyvraz_Redner92}.
The particles perform random walks,
subject to an infinite on-site repulsion between the same species
and $AB$ pairs annihilate  when they meet each other.
A random initial configuration with equal initial densities,
$\rho_A(0)=\rho_B(0)$ (a nontilted surface), decays to the absorbing  OF
``empty" state.

Neglecting spatial correlations gives rise to the rate equation
\begin{equation}
\frac{d\rho_A(t)}{dt} = \frac{d\rho_B(t)}{dt} = - \rho_A(t)\rho_B(t) ,
\end{equation}
with  $\rho(t) \sim t^{-1}$ as solution.
In the absence of on-site repulsion between the particles,
it has been shown rigorously that
this mean-field behavior, $\rho \sim t^{-1}$, holds
in dimensions higher than 4 and that the particle density
decays as $\rho(t) \sim t^{-\theta}$ with
$\theta = d/4$ in dimensions $d\leq 4$~\cite{Bramson_Lebowitz88}.
This contradicts the naive expectation that $\theta$ be equal to the
inverse of the random walk dynamic exponent $\theta =1/z_{\rm{rw}}=1/2$.

The following scaling argument explains this result
intuitively~\cite{Leyvraz_Redner92}.
Let ${\cal D}_R$ be the difference in the number of $A$ and $B$ particles
in a cube of size $R^d$.
For random (uncorrelated) initial configurations,
${\cal D}_R$ will be of order $\sqrt{\rho(0) R^d}$.
As the system evolves, each particle diffuses over a distance
$\sim t^{1/z_{\rm{rw}}}$ during time $t$
(ignoring on-site repulsion).
So after time $t_R \sim R^{z_{\rm{rw}}}$,
all members of the minority species found a partner
in the cube and have annihilated.
This leaves the region occupied by the majority species only.
Therefore, the particle density decays as
\begin{equation} \label{dens}
\rho(t) \sim {\cal D}_R /R^d|_{R\sim t^{1/z_{\rm{rw}}}}
\sim \sqrt{\rho(0)}\ t^{-d/2 z_{\rm rw}} .
\end{equation}
This argument gives the correct value of the decay exponent
$\theta = d/2z_{\rm{rw}}= d/4$ for $d\leq 4$.
Several numerical simulations~\cite{Toussaint_Wilczek83,Kang_Redner84}
found that the
on-site repulsion between the same species does not alter the decay
exponent $\theta$.

The same argument can be extended to the time evolution
of the pair densities and the interparticle distances~\cite{Leyvraz_Redner92}.
We need  them later in this section.
The size of a domain occupied by one single species of particles
(e.g., a train of $A$'s), $l_D$,
grows in time with the same exponent as the random walk radius,
$l_D \sim t^{1/z_{\rm rw}}$.
The $AB$ pair density scales in $d=1$ as $\rho_{AB}\sim l_D^{-1}$
because each domain with a single species of particles is bounded by two
$AB$ pairs.
So $\rho_{AB}$ decays with $\theta_{AB}=1/z_{\rm{rw}}$, i.e., decays faster
than the
particle density $\rho$; $\theta_{AB}= 2 \theta$.

The $AA$ and $BB$ pair densities scale differently.
They  are equal for nontilted initial states.
The pair densities add up to the total particle density,
$\rho = \rho_{AB}+\rho_{AA}+\rho_{BB}$ (see section 2).
Since  $\rho_{AB}(t)$ decays faster than $\rho(t)$,
$\rho_{AA}=\rho_{BB}$ must have the same asymptotic behavior as the
particle density $\rho$,
i.e.,~$\theta_{AA} = \theta = 1/2z_{\rm{rw}}$ in one dimension.

\begin{figure}
\centerline{\epsfxsize=75mm \epsfbox{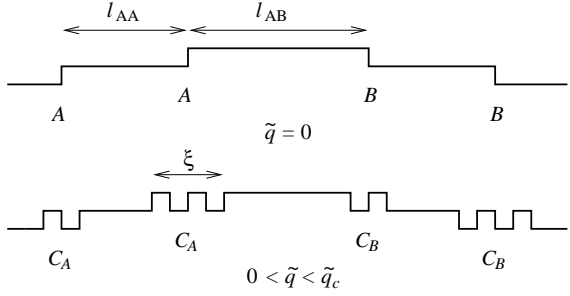}}
\caption{Schematic typical surface configurations at the absorbing side
of the DI transition, $0<\tilde q<\tilde q_c$.}
\label{C_A&C_B}
\end{figure}

The interparticle distances diverge in the long time limit.
Define $l_{AA}$ ($l_{AB}$) as the average distance between the
nearest neighboring particles of the same (different) species (see
Fig.~\ref{C_A&C_B}).
These distances grow via pair annihilations of $AB$ pairs.
The $A$ and $B$ particle meet through random walk fluctuations
at time intervals of order $l_{AB}^{z_{\rm{rw}}}$.
Therefore, the particle density decays as
$d\rho(t)/dt \sim - \rho_{AB}(t) / l_{AB}^{z_{\rm{rw}}}$,
which leads to $l_{AB}\sim t^{(2z_{\rm{rw}}-1)/2z_{\rm{rw}}^2} \sim t^{3/8}$.
The sum of all interparticle distances adds up to the size of the system:
\begin{equation}
\label{length_sum}
(\rho_{AA}+\rho_{BB}) l_{AA} + \rho_{AB} l_{AB}\simeq 1.
\end{equation}
The second term becomes negligible in the asymptotic limit, which yields
$l_{AA} \sim \rho_{AA}^{-1} \sim t^{1/2z_{\rm{rw}}}$.  $l_{AB}$ diverges
faster than
$l_{AA}$ but still slower than the interparticle distance of the ordinary
random walk
problem.

In summary, the dynamics at point $S$
belongs to the random walk (diffusion) universality class.
Starting with the nontilted random initial configurations,
the density of $AB$ pairs decays  with the naive
random walk dynamic exponent, $\theta_{AB}=1/z_{\rm{rw}}$, but
the density of $AA$ pairs and the particle density
decay much slower, with $\theta_{AA}=\theta=1/2z_{\rm{rw}}$.

The above results assume that the initial condition is random.
The exponents change when we modify the initial state.
The factor 2 appearing in the exponents $\theta_{AA}$ and $\theta$
originates from the random nature of the rough
initial configurations discussed in the beginning of this subsection.
Suppose it were correlated
such that the initial value for the surplus of particles of one
species in a box of size  $R^d$ scales with a different power,
i.e.,~${\cal D}_R \sim [\rho(0) R^d]^x$.
That leaves $\theta_{AB}=1/z_{\rm{rw}}$ unchanged, but modifies the asymptotic
behavior of the particle density and $AA$ pair density
to $\theta=\theta_{AA}=(1-x)/z_{\rm{rw}}$ in one dimension.
The exponents for the interparticle distances change accordingly.
The nontilted random initial configurations correspond to $x=1/2$ and
the $AB$ ordered initial configurations to $x=0$. The tilted random initial
configurations should correspond to $x=1$.

For initial configurations inside DOF subspace,
$\rho_{AA}$ and $\rho_{BB}$ are always equal to zero.
Then the particle density becomes equivalent to $\rho_{AB}$ and
decays with the ``conventional" random walk dynamic exponent,
$\rho\sim t^{-1/z_{\rm{rw}}}$.

Let's now investigate how this dependence of the decay critical exponents
on the initial states generalizes along the entire $\tilde r =0$ line,
according to the same type of reasoning.

\subsection{the absorbing phase: $0<\tilde{q}<\tilde{q}_c$}

At the absorbing side of the DI transition point,
the asymptotic time scaling behavior remains
the same as at point $S$ except for one important detail.
Clouds of particles take over the role of single particles.
Consider an initial rough state with a low particle density.
Each of these $A$ and $B$ particles broadens itself quickly  into a small cloud
of particles with a characteristic width $\xi$ via branching processes
(see Fig.~\ref{C_A&C_B}).
These clouds are created by OPB type branching,
and therefore preserve $ABAB$ alternating local order.
This broadening is governed by the DI type dynamics, like in defect
dynamics with a single starting particle.
The width of the clouds $\xi$ is finite and of the order of the
DI correlation length.
These clouds  are well defined in the asymptotic limit
because the distances between them, the length scales, $l_{AA}$ and $l_{AB}$,
diverge in time while  $\xi$ remains finite.

There are two topologically distinct types of clouds:
the $C_A$ clouds,  nucleated from a single $A$ particle and
with $A$'s at both  edges ($AB\cdots BA$), and
the $C_B$  clouds ($BA\cdots AB$).
Clouds diffuse through hopping, branching, and pair annihilation of
bare particles.
$C_AC_B$ pairs can annihilate each other.
Clouds do not branch.
They could in principle, but a branching process like $C_A \to C_A C_B C_A$,
requires a collective sequence of microscopic events involving many particles,
and at length scales $l>\xi$ this does not happen.

The clouds, at the DI length scale $\xi$, obey therefore the same
dynamic rules as the diffusion-limited pair
annihilation process of bare particles at point $S$,
with renormalized probabilities.
The  density of the clouds, $\tilde{\rho}(t)$,
and the cloud pair densities,
$\tilde{\rho}_{AA}$ (of $C_A C_A$ cloud pairs)
and $\tilde{\rho}_{AB}$ (of $C_A  C_B$ pairs),
must scale in the same way as particle densities at point $S$,
i.e.,~$\tilde{\rho}(t) \sim t^{-1/2z_{\rm{rw}}}$,
$\tilde{\rho}_{AA}(t) \sim  t^{-1/2z_{\rm rw}}$, and
$\tilde{\rho}_{AB}(t) \sim t^{-1/z_{\rm{rw}}}$.

In numerical simulations we measure the bare particle densities.
Those are related to the cloud densities as follows.
Each $C_A C_A$ pair of clouds contains  only one
$A A $ pair of bare particles, since each cloud consists out of a perfectly
ordered $ABAB$ train of particles.
This implies that $\rho_{AA} = \tilde{\rho}_{AA}$ and
$\theta_{AA} = 1/2 z_{\rm{rw}}$.
The number of $AB$ pairs of bare particles in each cloud
is proportional to its width $\xi$, so $\rho_{AB} \simeq \xi \tilde{\rho}(t)$
and $\theta_{AB} = 1/2z_{\rm{rw}}$.
The bare particle density is equal to the sum of all pair densities,
$ \rho =\rho_{AA}+\rho_{BB}+\rho_{AB}$.
Therefore $\rho$ scales with the slowest exponent,
i.e., $\theta = 1/2z_{\rm{rw}}$.

The only difference with point $S$ is that
all three densities decay with the same modified random walk exponent,
$\sim  t^{-1/2z_{\rm rw}}$.
The random walk nature of the dynamics is completely obscured now.
The random walk nature of the decay  dynamics
manifests itself only inside the subspace with
$ABAB$ ordered initial configurations; there, $\theta = 1/z_{\rm {rw}}$.

\subsection{the DOF active phase: $\tilde{q}>\tilde{q}_c$}

\begin{figure}
\centerline{\epsfxsize=75mm \epsfbox{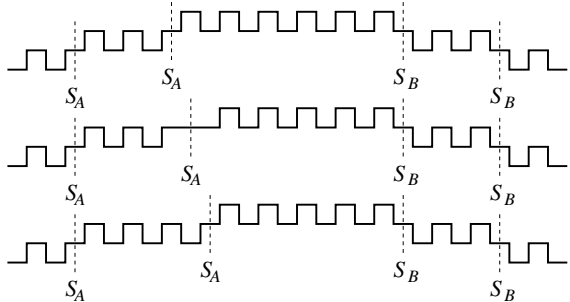}}
\caption{Schematic surface  configurations at $q=1$.
They illustrate that hopping of the solitons
requires a sequence of two branching processes.}
\label{soliton}
\end{figure}

In the active phase, solitons play the same role as the particle clouds
do at the absorbing side of the DI phase transition.
First, consider the limiting case $\tilde{q}=\infty$~($p=0,q=1$)
where the hopping probability becomes negligible with respect to
OPB type branching.
Any random initial configuration develops quickly by OPB type branching
into fully-developed DOF domains separated by nearest-neighbor pairs
of $AA$ (step-up step-up) or $BB$ (step-down step-down)
particles as shown in Fig.~\ref{soliton}.
These  $AA$ and $BB$ pairs denoted by ${S}_{A}$ and
${S}_{B}$ in Fig.~\ref{soliton} are the topological excitations
against the DOF phase and will be called the A- and B-type solitons.
The density of each soliton type is equal to the pair densities of
bare particles, $\rho_{AA}$ and $\rho_{BB}$, respectively.

At  $\tilde{q}\to\infty$, all activity is blocked inside each DOF domain,
because any attempt of OPB type branching is rejected due to
the infinite on-site repulsion between the same species
and the fact that  the DOF structure is close-packed.
Only the solitons at the boundaries of the DOF domains are active.
Their dynamics is  basically identical
to that of the bare particles at  point $S$.
Soliton diffusion is a second-order OPB process.
Each soliton can hop to a next-nearest-neighbor site
by applying OPB type branching twice~(Fig.~\ref{soliton}).
The solitons of the same species repel each other
and ${S}_{A} {S}_{B}$ soliton pairs can annihilate each other
when they meet. There is no mechanism to create solitons.

At finite $\tilde{q}$, the solitons broaden.
Their width $\xi$ is the DI correlation length in the active phase.
Just as the clouds in the absorbing phase, these broadened solitons
must obey effective dynamics at length scales $l>\xi$ identical to
those of the sharp solitons at $\tilde{q}\to\infty$; i.e., identical to the
particles at point $S$.
The soliton density decays with the same exponent $\theta$ as
that of the clouds at the other side of the transition.

Since the soliton density is equal to the particle pair densities
$\rho_{AA}$ and $\rho_{BB}$, we obtain $\rho_{AA} \sim t^{-\theta_{AA}}$
with $\theta_{AA}=1/2z_{\rm rw}$.
The particle density is finite in the steady active state.
So $\rho$ and $\rho_{AB}$ do not decay as power laws, but  remain nonzero.

\subsection{at the critical point: $\tilde{q}=\tilde{q}_c$}

The decay dynamics at the DI critical point can be discussed
equally well from the cloud or the soliton perspective
(the absorbing state or the active phase point of view).
At the critical point the DI correlation length diverges in time as
$\xi \sim t^{1/z_{\rm{DI}}}$
with the DI dynamic exponent $z_{\rm{DI}}=\nu_\parallel/\nu_\perp\simeq 1.76$.
This length diverges much faster than the typical distance between
clouds ($l_{C_A C_A}\sim t^{1/4}$ and $l_{C_A C_B}\sim t^{3/8}$)
(or solitons from the other point of view).
This implies that the motions of the clouds (solitons) become correlated
by DI critical fluctuations.
Their diffusion is not driven by random walks with
dynamic exponent $z_{\rm{rw}}=2$, but by correlated random walks
with the DI dynamic exponent $z_{\rm{DI}}$.
We expect this to be the only change in the scaling theory for
the clouds (solitons).
That means that we only need to replace $z_{\rm{rw}}$ by $z_{\rm{DI}}$.
The total density of clouds should scale as
$\tilde{\rho}(t) \sim \ t^{-1/2z_{\rm{DI}}}$,
the density of $C_A C_A$ cloud pairs  as
$\tilde{\rho}_{AA}(t) \sim \ t^{-1/2z_{\rm{DI}}}$,
and the density of $C_A C_B$ cloud pairs as
$\tilde{\rho}_{AB}(t)  \sim t^{-1/z_{\rm{DI}}}$.
The densities of the solitons (on the opposite
side of the transition) scale identically.

Next, we need to establish  how these cloud and soliton densities are related
to the bare particle densities.
We did this for both, and the answer is the same,
but the argument is easier from the clouds perspective
and therefore we present only the former.

The number of $AA$ particle pairs
is equal to the number of $C_A C_A$ soliton pairs
(like above in subsection B).
Therefore the $AA$ particle and $BB$ particle pair densities
scale as $\rho_{AA (t)}=\rho_{BB}(t) \sim t^{-1/2z_{\rm{DI}}}$.
The total density of clouds scales as
$\tilde{\rho}(t) \sim \ t^{-1/2z_{\rm{DI}}}$,
and therefore the average distance between diverges as
$\tilde{l}_D(t) \sim \ t^{1/2z_{\rm{DI}}}$.
To find the total density of particles, we need to know
how many clouds there are and how many particles each cloud carries.
The width of the clouds, $l_w$, diverges.
For a single isolated cloud this happens with the DI correlation length,
$\xi (t) \sim t^{z_{\rm{DI}}}$.
This is much faster than the intercloud distance, $\tilde{l}_D$.
Therefore the width of the clouds is limited by the latter, $l_w=\tilde{l}_D$.
The density of particles inside  each cloud scales as
$\rho_p \sim l_w^{-\beta/\nu_\perp}$ (Eq.~(\ref{rho_s})),
and their total number  therefore as
$N_p \sim l_w^{1-\beta/\nu_\perp} \sim t^{(1-\beta/\nu_\perp)/2z_{\rm{DI}}}$.
The total density of steps scales as
$\rho\sim N_p \tilde \rho$.
Putting all this together gives
$\rho \sim t^{-\beta/(2\nu_\perp z_{\rm{DI}})}\sim t^{-\beta/2\nu_{||}}$~\cite{explain3}.

The density of $\rho_{AB}$ pairs is related to the other two by the relation
$\rho=2\rho_{AA}+ \rho_{AB}$ and therefore  must scale with the slowest
power law,
$\rho_{AB} \sim t^{-\beta/2\nu_{||}}$.

The density exponents are smaller by a factor of 2, compared
to their asymptotic behavior starting from the $AB$ ordered initial
configurations (see section 3).
This is the same factor of 2 found at point $S$
and everywhere else along the  $\tilde r=0$ line.
This factor reflects the random roughness of the initial configurations.
It changes for correlated initial rough configurations
in the same manner as discussed in subsection A for the $S$ point,
i.e.,~ simply replace ${1\over2} \to (1-x)$.

In conclusion, the above arguments predict that
at the PR transition point the
step density scales as $\rho \sim t^{-(1-x)\beta/\nu_{||}}$,
that the $AB$ step pair density  scales with the same exponent,
and that the  $AA$ and $BB$ step densities scale as
$\rho_{AA}=\rho_{BB} \sim t^{-(1-x)/z_{\rm{DI}}}$.
$z_{\rm{DI}}$, $\beta$, and $\nu_{||}$ are directed Ising critical exponents,
but $x$ represents the correlations in the initial configuration.
The initial state  properties persist into the asymptotic scaling properties.

\section{Numerical simulations for the density decay}\label{sec5}

The scaling theory in the previous section is heuristic,
and certainly not exact.
It is somewhat questionable in particular at the PR transition
because we assume that
the clouds (and solitons) remain valid concepts,
while their widths actually
like to diverge faster than allowed by the inter cloud (soliton) distances.

To test these predictions, we perform Monte Carlo simulations
starting from a random initial state where
particles are distributed randomly on a lattice of size
$L=2^{15}$ with initial densities $\rho_A=\rho_B=1/4$.
We apply periodic boundary conditions.
The time evolutions of the  densities $\rho(t)$, $\rho_{AB}(t)$, and
$\rho_{AA}(t)=\rho_{BB}(t)$, are monitored
up to $t = 10^4$ and averaged over $100$ independent runs.
A few  simulations on a larger lattice of size $L=2^{16}$
demonstrate  that $L=2^{15}$ is adequate to describe the scaling
behavior up to time  $t= 10^4$.

\begin{figure}
\centerline{\epsfxsize=85mm \epsfbox{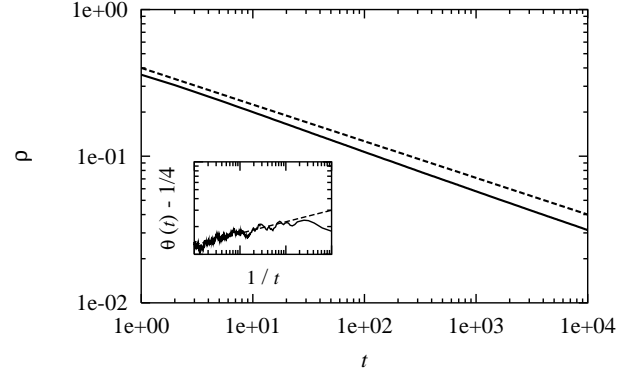}}
\caption{The particle density $\rho$ vs. $t$ at point $S$.
The dashed line has the expected slope $-1/4$.
The inset shows that the effective exponent
$\theta(t)$ approaches $1/4$ with a power law correction,
i.e.,~$\theta(t)-1/4 \sim t^{-1/8}$.
The dashed line in the inset has the predicted slope $1/8$.}
\label{Spoint}
\end{figure}

First, we test the density decay at point  $S$  ($\tilde{q}=\tilde{r}=0$).
In the absence of the on-site repulsion between the same species,
the total particle density $\rho$ should decay algebraically with exponent
$\theta=1/4$~\cite{Bramson_Lebowitz88}.
It is important to confirm explicitly
that the infinite on-site repulsion between the same species
in our model does not change this result.
In Fig.~\ref{Spoint}, $\rho$ is plotted against $t$
on a log-log scale  (the solid line).
It seems that the density decays slightly faster than $t^{-1/4}$
(the dashed line).
Similar results were found previously~\cite{Leyvraz_Redner92}.
This is a correction-to-scaling effect.
Insert the leading scaling behaviors of
$l_{AB}\sim t^{3/8}$ and $\rho_{AB}
\sim t^{-1/2}$ into Eq.~(\ref{length_sum}).
This gives  $2 \rho_{AA}\times l_{AA} \simeq (1 - a t^{-1/8})$,
and shows the presence of a generic ${\cal O}(t^{-1/8})$
type correction-to-scaling term.
To isolate this term,
we define an effective exponent
$\theta(t) \equiv -\ln[\rho(t)/\rho(t/m)]/\ln m$ with $m=8$.
The leading scaling exponent $\theta$ is given by the
limiting value of $\lim_{t\rightarrow\infty}\theta(t)$ and the
correction-to-scaling behavior is contained in $\theta(t)-\theta$.
The log-log plot of $\theta(t)-1/4$ against $1/t$
is shown in the inset of Fig.~\ref{Spoint}.
We find that $\theta(t)-1/4$ scales clearly  as $t^{-1/8}$.
This confirms that $\rho(t) \sim t^{-1/4} ( 1+{\cal O}(t^{-1/8}))$.
This correction-to-scaling term decays very slowly.
Fitting $\rho(t)$ to a simple power-law form therefore fails to
produce the correct value of the leading exponent.
We measured also the effective exponents $\theta_{AA}(t)$ and
$\theta_{AB}(t)$ at point $S$.
They approach $1/4$ and $1/2$, respectively,
with  power-law corrections to scaling as well.

The step density suffers from the same type of corrections to scaling
everywhere along the $\tilde r =0$ line.
Log-log plots of the density versus time are not straight lines.
They  curve a little.
We analyze the data in the following manner.
We construct estimates
$\theta^{(k)}$~($k=1,2,3$) for the exponent $\theta$ by fitting
the measured density $\rho(t)$ to a power law in the time interval
$10^k\le t <10^{k+1}$~($k=1,2,3$).
Approximants  for $\theta_{AA}^{(k)}$ and $\theta_{AB}^{(k)}$ are
constructed in the same way.
As time increases the correction-to-scaling term contributes
less and less,
and the estimates should converge to the correct values of the
leading decay exponents.

\begin{figure}
\centerline{\epsfxsize=85mm \epsfbox{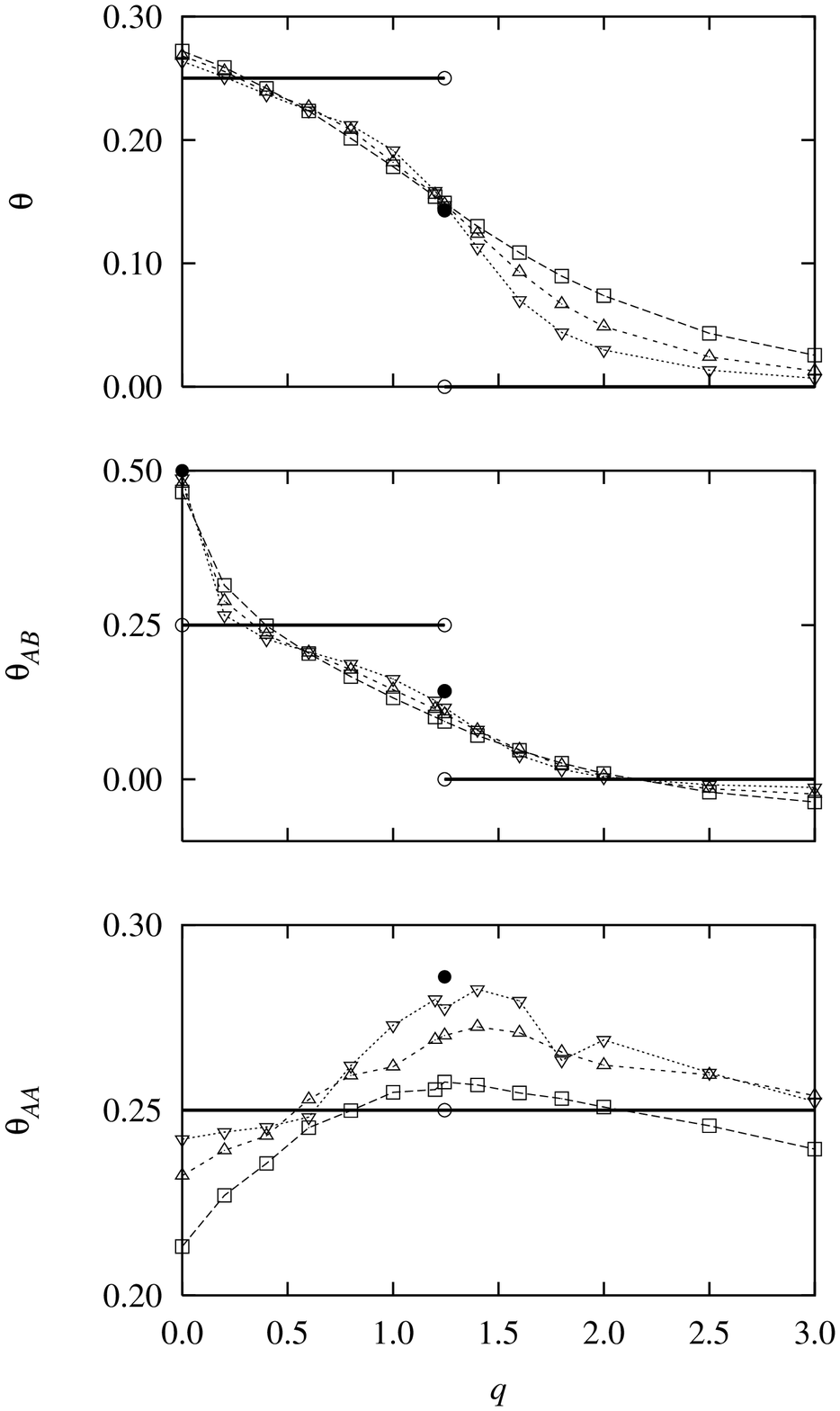}}
\caption{
Estimates for the density exponents.
The squares~($\Box$),
up-triangles~($\bigtriangleup$), and down-triangles~($\bigtriangledown$)
correspond to $k=1,2,$ and 3, respectively.
The predicted values of the exponents
are represented by solid lines and circles.
}
\label{theta}
\end{figure}

In Fig.~\ref{theta}, the estimates are presented and compared to
the predicted values from the scaling theory in the previous section.
The step density exponent $\theta$
is predicted to take the value of $\theta=1/4$ below the transition and
$\frac{1}{2}(\beta/\nu_\parallel)\simeq 0.141$ at the critical point.
Above the transition, the density saturates to a finite value,
i.e.,~$\theta=0$.
The estimates for $\theta$ merge to the predicted value
at the critical point $\tilde{q}_c=1.245$.
Above the transition, $\theta^{(k)}$ become smaller as $k$ increases, which
is consistent with  $\theta=0$.
Below the transition, the convergence is slow
(due to  strong  corrections to scaling) but
compatible with $\theta=1/4$.

The exponent $\theta_{AB}$ is predicted to take the value of
$\theta_{AB}=1/2$ at point $S$,
$\theta_{AB}=1/4$ along  $0<\tilde{q}<\tilde{q}_c$,
$\theta_{AB}=\frac{1}{2}(\beta/\nu_\parallel)\simeq 0.141$ at
the transition point, and $\theta_{AB}=0$ in the DOF phase.
The estimates, shown in Fig.~\ref{theta}, converge not as well as
those for $\theta$, but are still compatible with the theoretical predictions.
The exponent $\theta_{AA}$ is expected to take the value of
$\theta_{AA}=1/2z_{\rm{DI}}\simeq 0.284$
at the critical point and $\theta_{AA}=1/4$ everywhere else.
The data for this exponent converge slower than the other two,
but they are compatible with the theoretical results as well.

\section{Conclusions }\label{sec6}

In this paper we presented a surface physics application of
dynamic phase transitions in the directed Ising universality class.
We introduced a model for the relaxation of a one-dimensional  interface
starting from arbitrary initial (rough) configurations,
e.g., the straightening  of a step on a vicinal surface.
This model undergoes a  DI type dynamic
preroughening transition between a perfectly ordered flat (OF)
(absorbing) stationary state and a
disordered flat (DOF) (active) stationary phase.
The step becomes perfectly straight or
straight in average with randomly placed kinks but in perfect
up-down alternating order.

The OF and DOF phases are both unstable with respect to the
OBB type branching processes that break the up-down alternating order.
There we find a rough stationary state with
Edwards-Wilkinson type scaling behaviour.
The crossover exponent into the rough phase is determined numerically.

The asymptotic long time behaviour of the kink density
depends strongly on the statistical properties and correlations
of the  initial configurations.
Information about the correlations in the rough initial state
(controllable experimentally by sputtering for example) never gets lost.
It obscures the random walk nature in the absorbing phase
and the DI nature at criticality.
We develop a scaling theory for the decay dynamics of the
various kink densities and predict the values of the decay
exponents. Numerical simulations confirm these.

It is noteworthy to mention
a different recent application of absorbing phase transitions to
one-dimensional interface problems by Alon {\em et al.}~\cite{Alon_etal96},
which describes the dynamic roughening
phase transition from a smooth phase into a rough phase of
Kardar-Parisi-Zhang type~\cite{KPZ}.
They considered
a solid-on-solid type model where
particles can adsorb at any site but desorption takes
place only at existing steps.
In other words, one can build mountains but is not allowed to dig new holes.
This leads to the dynamic roughening phase transition
at a finite value of the adsorption rate, which is triggered by
the absorbing nature of the lowest level. This phase transition
belongs to the DP universality class.
It may be interesting to introduce in our model a symmetry-breaking
field between adsorption and desorption processes like in the above model.
Generalization of our model in this direction is currently under study.

\acknowledgements
This work was supported in part by NSF grant DMR-9700430,
by the Korea Research Foundation (JDN),
and by the academic research fund of the Ministry of
Education, Republic of Korea (Grant No.~97-2409) (HP).

\narrowtext


\end{multicols}

\begin{references}
\bibitem{explain0} A smooth (but not perfectly flat) phase was found stable
against the roughening degree of freedom in other models~\cite{Alon_etal96}.
But this is not a true absorbing phase where the dynamics are completely dead.
\bibitem{explain1} The critical point of the BAW model with dynamic branching
is at $\tilde q \simeq 0.959$~\cite{Kwon_Park95}. In our model, some
branching processes
are rejected as explained in section 3. So the absorbing phase in our model
becomes a little bit larger than in the BAW model.
\bibitem{explain2} This is the conventional definition of the dynamic
exponent $z$.
We do not take the traditional usage of $z$ as the exponent of the mean-square
distance of
spreading at the absorbing transition. This
spreading exponent can be expressed as $2/z$ in terms
of
the dynamic exponent $z$.
\bibitem{explain3} {
The distances between nearest neighbour clouds of the same species
($\tilde{l}_{AA}$) and opposite species ($\tilde{l}_{AB}$)
diverge with different exponents.
The divergence of the cloud widths
is limited by these distances, and therefore
depends on their local neighbourhoods.
It suffices to deal with the average distance $\tilde{l}_w$,
as done in the text,
since $\tilde{l}_{AA}\simeq \tilde{l}_w$ is the slowest
of the two and turns out to dominate.}
\end{references}
\end{document}